\documentclass[12pt]{article}
\usepackage{amsmath}
\usepackage{amssymb}
\usepackage{latexsym}

\let\textquotedbl="

\begin{document}

\title{Another derivation of generalized Langevin equations}
\author{R. Dengler}
\maketitle

\begin{abstract}
The formal derivation of Langevin equations (and, equivalently Fokker-Planck equations)
with projection operator techniques of Mori, Zwanzig, Kawasaki and others apparently not has
widely found its way into textbooks. It has been reproduced dozens of times on the fly with many references to
the literature and without adding much substantially new.
Here we follow the tradition, but strive to produce a self-contained text.
Furthermore, we address questions that naturally arise in the derivation.
Among other things the meaning of the divergence of the Poisson brackets
is explained, and the role of nonlinear damping coefficients is clarified.
The derivation relies on classical mechanics, and encompasses everything one can construct from point particles and potentials:
solids, liquids, liquid crystals, conductors, polymers, systems with spin-like degrees of freedom ... Einstein relations
and Onsager reciprocity relations come for free.
\end{abstract}

\section{Notation}

Coordinates in phase space are denoted by $\Gamma=\left\{ p_{n},q_{n}\right\} $,
where $p$ are the momenta and $q$ the coordinates. A Hamiltonian
$H\left(\Gamma\right)$ defines a flow in phase space according to
\begin{eqnarray*}
d\Gamma/dt & = & \left[\Gamma,H\right]=iL\Gamma,
\end{eqnarray*}
where $\left[X,Y\right]=\sum\left(\frac{\partial X}{\partial q}\frac{\partial Y}{\partial p}-\frac{\partial X}{\partial p}\frac{\partial Y}{\partial q}\right)$
is the Poisson bracket and $L=i\left[H,...\right]$ is the Liouville
operator. The expression
\[
\Gamma\left(\Gamma,t\right)=e^{iLt}\Gamma
\]
is a function in phase space mapping every point $\Gamma$ to the
value of the point a time step $t$ in the future. A simple example
is $H=\left(p^{2}+q^{2}\right)/2$ with $\Gamma\left(\Gamma,t\right)=\left(p\left(t\right),q\left(t\right)\right)=\left(p\cos t-q\sin t,\:q\cos t+p\sin t\right).$
More generally $A\left(\Gamma,t\right)=e^{iLt}A\left(\Gamma\right)$
is the value of an arbitrary function $A\left(\Gamma\right)$ at time
$t$ when the system is in state $\Gamma$ at $t=0$. The density
of the phase space points of an ensemble is defined as
\[
\rho\left(\Gamma,t\right)=\left\langle \delta\left(\Gamma-\Gamma\left(\Gamma_{0},t\right)\right)\right\rangle _{\Gamma_{0}}=\left\langle \prod\delta\left(p_{n}-p_{n}\left(t\right)\right)\delta\left(q_{n}-q_{n}\left(t\right)\right)\right\rangle _{\left\{ p_{n}\left(0\right),q_{n}\left(0\right)\right\}},
\]
where the angular bracket denotes the ensemble average. The time derivative of $\rho$ at a given point $\Gamma$ is
\[
\frac{\partial}{\partial t}\varrho\left(\Gamma,t\right)=\frac{\partial\rho}{\partial p}\left(-\dot{p}\right)+\frac{\partial\rho}{\partial q}\left(-\dot{q}\right)=\frac{\partial\rho}{\partial p}\frac{\partial H}{\partial q}-\frac{\partial\rho}{\partial q}\frac{\partial H}{\partial p}=\left[H,\varrho\right]=-iL\rho.
\]
A formal solution of this equation is $\varrho\left(\Gamma,t\right)=e^{-iLt}\varrho\left(\Gamma,0\right)$.
Of particular interest is the probability
\[
p\left(a,t\right)=\int d\Gamma\delta\left(a-A\left(\Gamma\right)\right)\varrho\left(\Gamma,t\right)=\int d\Gamma\delta\left(a-A\left(\Gamma,t\right)\right)\varrho\left(\Gamma\right)
\]
to observe values $a=\left\{ a_{i}\right\} $ of observables $A=\left\{ A_{i}\right\} $
at time $t$ when the ensemble is in a state with phase space density
$\varrho\left(\Gamma\right)$ at time $t=0$. The equilibrium distribution function of
observables is written in the form
\begin{equation}
p_{0}\left(a\right)da=const\times e^{-\beta H_{eff}\left(a\right)}da \label{eq:EffHamiltonian}
\end{equation}
where $H_{eff}\left(a\right)$ is called effective Hamiltonian.

\section{Projection Operator}

The Zwanzig projection\cite{Zwan1961} operator operates on functions in the $6$-$N$-dimensional
phase space of $N$ point partices and projects onto the
linear subspace of \textquotedbl{}slow\textquotedbl{} \emph{phase
space} functions. It was introduced by R. Zwanzig to derive a generic
master equation. It is mostly used in this or similar context in a
formal way to derive equations of motion for some \textquotedbl{}slow\textquotedbl{}
collective variables. A special subset of these functions is an enumerable
set of \textquotedbl{}slow variables\textquotedbl{} $A\left(\Gamma\right)=\left\{ A_{n}\left(\Gamma\right)\right\} $.
Candidates for some of these variables might be the long-wavelenght
Fourier components of conserved quantities like energy, momentum and mass density. Relaxation of such quantities
requires a transport of the conserved quantity over the distance of
a wavelength, and thus is much slower than the relaxation of local
quantities, which typically occurs within a few collision times. Another
example is the order parameter of a second order phase transition,
the dynamics of which slows down at the critical point. The
Zwanzig projection operator relies on these functions but doesn't
tell how to find the slow variables of a given Hamiltonian $H\left(\Gamma\right)$.

\subsection{Slow variables and scalar product}

A scalar product between two phase space functions $f_{1}\left(\Gamma\right)$
and $f_{2}\left(\Gamma\right)$ is defined by the equilibrium correlation
function\cite{Mori1965}
\begin{equation}
\Bigl\langle f_{1}f_{2}\Bigr\rangle_{0}=\int d\Gamma\rho_{0}\left(\Gamma\right)f_{1}\left(\Gamma\right)f_{2}\left(\Gamma\right),\label{eq:ExpValue_0}
\end{equation}
where $\rho_{0}\left(\Gamma\right)=Z^{-1}e^{-\beta H\left(\Gamma\right)}$
is the canonical equilibrium distribution. \textquotedbl{}Fast\textquotedbl{}
variables, by definition, are orthogonal to all functions $G\left(A\left(\Gamma\right)\right)$
of $A\left(\Gamma\right)$ under this scalar product. This definition
states that fluctuations of fast and slow variables are uncorrelated,
and according to the ergodic hypothesis this also is true for time
averages. If a generic function $f\left(\Gamma\right)$ is correlated
with some slow variables, then one can subtract functions of slow
variables until there remains the uncorrelated fast part of $f\left(\Gamma\right)$.
The definition also means that the average of a fast variable vanishes.
The product of a slow and a fast variable is a fast variable.

The concise notation
\[
\Bigl\langle f\Bigr\rangle_{0}=\int d\Gamma'\rho_{0}\left(\Gamma'\right)f\left(\Gamma'\right)
\]
which completely hides the integration variables is used systematically below.

\subsection{The projection operator}

Consider the continuous set of functions $\Phi_{a}\left(A\left(\Gamma\right)\right)=\delta\left(A\left(\Gamma\right)-a\right)=\prod\nolimits _{n}\delta\left(A_{n}\left(\Gamma\right)-a_{n}\right)$
with $a=\left\{ a_{n}\right\} $ constant. Any phase space function
$G\left(A\left(\Gamma\right)\right)$ depending on $\Gamma$ only
through $A\left(\Gamma\right)$ is a function of the $\Phi_{a}$,
namely
\[
G(A\left(\Gamma\right))=\int daG\left(a\right)\delta\left(A\left(\Gamma\right)-a\right).
\]
A generic phase space function $f\left(\Gamma\right)$ decomposes
according to
\[
f\left(\Gamma\right)=F\left(A\left(\Gamma\right)\right)+R\left(\Gamma\right)\text{,}
\]
where $R\left(\Gamma\right)$ is the fast part of $f\left(\Gamma\right)$.
To get an expression for the slow part $F\left(A\left(\Gamma\right)\right)$
of $f$ take the scalar product (\ref{eq:ExpValue_0}) with the slow
function $\delta\left(A\left(\Gamma\right)-a\right)$,
\begin{eqnarray*}
\left\langle \delta\left(A-a\right)f\right\rangle _{0} & = & \left\langle \delta\left(A-a\right)F\left(A\right)\right\rangle _{0}=F\left(a\right)\left\langle \delta\left(A-a\right)\right\rangle _{0}.
\end{eqnarray*}
This gives an expression for $F\left(a\right)$, and thus for the
operator $P$ projecting an arbitrary function $f\left(\Gamma\right)$
to its \textquotedbl{}slow\textquotedbl{} part depending on $\Gamma$
only through $A\left(\Gamma\right)$,
\begin{equation}
P\cdot f\left(\Gamma\right)=F\left(A\left(\Gamma\right)\right)=\frac{\left\langle \delta\left(A-A\left(\Gamma\right)\right)f\right\rangle _{0}}{\left\langle \delta\left(A-A\left(\Gamma\right)\right)\right\rangle _{0}}\text{.}\label{eq:Projection1}
\end{equation}
This expression agrees with the expression given by Zwanzig\cite{Zwan1961}, except
that Zwanzig uses a microcanonical ensemble and subsumes $H\left(\Gamma\right)$
under the slow variables.\footnote{A heat bath in contact with the boundary of the system is irrelevant for the dynamics in the inner parts
of the system. The only difference between canonical and microcanonical ensemble is that the microcanonical ensemble
doesn't have the $k=0$ degree of freedom of the energy density.}
The Zwanzig projection operator fulfills $P\cdot G\left(A\left(\Gamma\right)\right)=G\left(A\left(\Gamma\right)\right)$
and $P^{2}=P$. The fast part of $f\left(\Gamma\right)$ is $\left(1-P\right)f\left(\Gamma\right)$.
Functions of slow variables and in particular products of slow variables
are slow variables. The space of slow variables thus is an algebra.
The algebra in general is not closed under the Poisson bracket, including
the Poisson bracket with the Hamiltonian. The projection operator (\ref{eq:Projection1}) can also
be written in the form 
\begin{eqnarray}
P\cdot f\left(\Gamma\right) & = & \int\frac{da}{p_{0}\left(a\right)}\delta\left(a-A\left(\Gamma\right)\right)\Bigl\langle\delta\left(a-A\right)f\Bigr\rangle_{0}\label{eq:Projection2}
\end{eqnarray}
where $p_{0}\left(a\right)=\left\langle \delta\left(a-A\right)\right\rangle _{0}$
is the equilibrium distribution (\ref{eq:EffHamiltonian}) of $A$.

\section{Exact generalized Langevin Equations}

The starting point for the standard derivation of a Langevin equation
is the identity $1=P+Q$, where $Q$ projects onto the fast subspace.
It is important in the following that the Liouville operator normally
doesn't commute with the Zwanzig projection operator - time evolution
mixes fast and slow variables, $\left[L,P\right]\neq0$. 

Consider discrete small time steps $\tau$ with evolution operator
$U\simeq1+i\tau L$, where $L$ is the Liouville operator. The goal
is to express $U^{n}$ in terms of $U^{k}P$ and $Q\left(UQ\right)^{m}$.
The motivation is that $U^{k}P$ is a functional of slow variables
and that $Q\left(UQ\right)^{m}$ generates expressions which are fast
variables at every time step. The expectation is that fast variables
isolated in this way can be represented by some model data, for instance
a Gaussian white noise. The decomposition is achieved by multiplying
$1=P+Q$ from the left with $U$, except for the last term, which
is multiplied with $U=PU+QU$. Iteration gives
\begin{eqnarray*}
1 & = & P+Q,\\
U & = & UP+PUQ+QUQ,\\
U^{2} & = & U^{2}P+UPUQ+PUQUQ+QUQUQ,\\
... & = & ...\\
U^{n} & = & U^{n}P+\sum_{m=1}^{n}U^{n-m}P\left(UQ\right)^{m}+Q\left(UQ\right)^{n}.
\end{eqnarray*}
The last line can also be proved by induction.\footnote{The factor $UQ$ is rather formal. It means projecting away the slow
part of the observable after each time step. Hamiltonian and the trajectory
of the system point in phase space are unchanged.} Now assume $U=1+itL/n$ and perform the limit $n\rightarrow\infty$.
This directly leads to the operator identity of Kawasaki\cite{Kaw1973}
\[
e^{itL}=e^{itL}P+i\int_{0}^{t}dse^{i\left(t-s\right)L}PLQe^{isLQ}+Qe^{itLQ}.
\]
A generalized exact Langevin equation is obtained by applying this
equation to $dA\left(\Gamma,t\right)/dt=e^{itL}\left(dA\left(\Gamma,t\right)/dt\right)_{t=0}$,
\begin{eqnarray}
\frac{d}{dt}A\left(\Gamma,t\right) & = & V+K+R,\nonumber \\
V\left(\Gamma,t\right) & = & e^{itL}P\dot{A}\left(\Gamma,0\right),\label{eq:LangevinStep0}\\
K\left(\Gamma,t\right) & = & i\int_{0}^{t}dse^{i\left(t-s\right)L}PLQe^{isLQ}\dot{A}\left(\Gamma,0\right)=i\int_{0}^{t}dse^{i\left(t-s\right)L}PLR\left(\Gamma,s\right),\nonumber \\
R\left(\Gamma,t\right) & = & Qe^{itLQ}\dot{A}\left(\Gamma,0\right).\nonumber 
\end{eqnarray}
The contribution $R$ is the fluctuating force. The $K$ term is the damping, it simplifies with the expression
for the fluctuating force. The contribution $V$ is called mode coupling. It corresponds to reversible modes like spin or plasma waves.
Mode coupling term $V$ and damping term $K$ are functionals of $A\left(\Gamma,t\right)$ and
$A\left(\Gamma,t-s\right)$ respectively and can be simplified considerably.

\subsection{Mode Coupling}

Inserting the expression (\ref{eq:Projection2}) for the Zwanzig projection into the
$V$-term from eq.(\ref{eq:LangevinStep0}) gives
\begin{equation}
V_{i}\left(\Gamma,t\right)=e^{itL}P\left[A,H\right]=\int\frac{da}{p_{0}\left(a\right)}\delta\left(a-A\left(\Gamma,t\right)\right)\Bigl\langle\left[A_{i},H\right]\delta\left(a-A\right)\Bigr\rangle_{0}.\label{eq:V_1}
\end{equation}
The expectation value $\left\langle ...\right\rangle _{0}$ simplifies
with $\left[A_{i},H\right]\rho_{0}=-\left(1/\beta\right)\left[A_{i},\rho_{0}\right]$
and partial integration to
\begin{eqnarray*}
\left\langle ...\right\rangle _{0}\left(a\right) & = & \frac{-1}{\beta}\int d\Gamma^{\prime}\left[A_{i},\rho_{0}\right]\delta\left(a-A\left(\Gamma^{\prime}\right)\right)=\frac{-1}{\beta}\Bigl\langle\left[\delta\left(a-A\right),A_{i}\right]\Bigr\rangle_{0}\\
 & = & \frac{-1}{\beta}\biggl\langle\sum_{j}\frac{d\delta\left(a-A\right)}{dA_{j}}\left[A_{j},A_{i}\right]\biggr\rangle_{0}\\
 & = & \frac{-1}{\beta}\sum_{j}\frac{d}{da}_{j}\Bigl\langle\delta\left(a-A\right)\left[A_{i},A_{j}\right]\Bigr\rangle_{0}=\frac{-1}{\beta}\sum_{j}\frac{d}{dA}_{j}P\left[A_{i},A_{j}\right]p_{0}\left(A\right)\Bigl|_{A=a}.
\end{eqnarray*}
Inserting $\left\langle ...\right\rangle _{0}$ into (\ref{eq:V_1})
and writing $p_{0}\left(a\right)=const\times e^{-\beta H_{eff}\left(a\right)}$
finally gives
\begin{eqnarray}
V_{i}\left(\Gamma,t\right) & = & \frac{-k_{B}T}{p_{0}\left(A\right)}\sum_{j}\frac{d}{dA}_{j}P\left[A_{i},A_{j}\right]p_{0}\left(A\right)\Bigl|_{A=A\left(\Gamma,t\right)}=\sum_{j}P\left[A_{i},A_{j}\right]\frac{dH_{eff}}{dA_{j}}\Bigl|_{A\left(\Gamma,t\right)}+D_{i},\nonumber\\
D_{i}\left(\Gamma,t\right) & = & -k_{B}T\sum\frac{d}{dA}_{j}P\left[A_{i},A_{j}\right]\Bigl|_{A=A\left(\Gamma,t\right)}.\label{eq:ModeCoupling_V}
\end{eqnarray}
The mode coupling (\ref{eq:ModeCoupling_V}) is a functional of $A\left(\Gamma,t\right)$,
and the Langevin equation (\ref{eq:LangevinStep0}) with only $V$ on the
r.h.s. agrees with the classical (reversible) equation of motion of variables $A$
except for the purely kinematic divergence $D\left(A\right)$. In
fact, $D$ actually vanishes in most standard cases and it is reasonable
to suggest to use variables $A$ for which $D=0$. The meaning of
this condition is explained in detail in the appendix.

\subsection{Damping term}

Inserting the expression (\ref{eq:Projection2}) for the projection operator into the
damping term $K_{i}$ from eq.(\ref{eq:LangevinStep0}) gives
\begin{eqnarray*}
K_{i} & = & \int_{0}^{t}ds\int\frac{da}{p_{0}\left(a\right)}\delta\left(A\left(t-s\right)-a\right)\Bigl\langle\delta\left(a-A\right)iLR_{i}\left(s\right)\Bigr\rangle_{0}.
\end{eqnarray*}
The Liouville operator in the expectation value $\left\langle ...\right\rangle _{0}$
can be moved to the $\delta$-function by means of a partial integration,
which gives the negative time derivative of the delta function,
\[
-\frac{d}{dt}\delta\left(a-A\right)=-\sum_{j}\frac{d\delta\left(a-A\right)}{dA_{j}}\dot{A}_{j}=\sum_{j}\frac{d}{da_{j}}\delta\left(a-A\right)\dot{A}_{j}
\]
and thus
\[
K_{i}=\int_{0}^{t}ds\int\frac{da}{p_{0}\left(a\right)}\delta\left(A\left(t-s\right)-a\right)\sum_{j}\frac{d}{da_{j}}\left\langle \delta\left(a-A\right)\dot{A}_{j}\left(0\right)R_{i}\left(s\right)\right\rangle _{0}.
\]
The product of the slow part of $\dot{A}_{j}$ and $\delta\left(a-A\right)$
is a slow variable and doesn't contribute to $\left\langle ...\right\rangle _{0}$.
It therefore can be replaced with $Q\dot{A}_{j}\left(0\right)=R_{j}\left(0\right)$
from eq.(\ref{eq:LangevinStep0}). There results the exact expression
\begin{equation}
K_{i}\left(\Gamma,t\right)=\sum_{j}\int_{0}^{t}ds\left(\frac{1}{p_{0}\left(a\right)}\frac{d}{da_{j}}p_{0}\left(a\right)\Lambda_{j,i}\left(a,s\right)\right)_{a=A\left(t-s\right)},\label{eq:K_Generic}
\end{equation}
where
\begin{equation}
\Lambda_{j,i}\left(a,s\right)=\frac{1}{p_{0}\left(a\right)}\Bigl\langle\delta\left(a-A\right)R_{j}\left(0\right)R_{i}\left(s\right)\Bigr\rangle_{0}=\Bigl\langle R_{j}\left(0\right)R_{i}\left(s\right);a\Bigr\rangle_{0}\label{eq:Lambda_Def}
\end{equation}
are the correlation functions of the fluctuating forces restricted
to a subspace with given values $a$ for the slow variables. $K\left(\Gamma,t\right)$
in effect is a functional of $A\left(t-s\right)$ and $\Lambda_{j,i}\left(A\left(t-s\right),s\right)$.
The exact evolution $A\left(t-s\right)$ of $A$ over the time interval
$\left[s,t\right]$ here is uninteresting, the nontrivial part is
the evolution $R\left(s\right)$ of the fluctuating force $R$ over
the time interval $\left[0,s\right]$. If the variables $A$ change
much more slowly than $R$ then one can think of $\Lambda_{j,i}\left(a,s\right)$
as a time average in a situation where the variables $A$ have the
value $a$.

\subsection{Fluctuating force}

The fluctuating force $R(t)=R\left(\Gamma,t\right)$ is a fast variable
according to definition (\ref{eq:LangevinStep0}) at every time $t$
and therefore $\left\langle R\left(t\right)\right\rangle _{0}=0$.
It also is easy to verify that $R$ is a stationary process: the correlation function
\begin{eqnarray*}
\Bigl\langle R_{i}\left(s\right)R_{j}\left(t+s\right)\Bigr\rangle_{0} & = & \Bigl\langle Qe^{isLQ}R_{i}\left(0\right)\cdot Qe^{i\left(t+s\right)LQ}R_{j}\left(0\right)\Bigr\rangle_{0}\\
 & = & \Bigl\langle R_{i}\left(0\right)\cdot Qe^{itLQ}R_{j}\left(0\right)\Bigr\rangle_{0}=\Bigl\langle R_{i}\left(0\right)R_{j}\left(t\right)\Bigr\rangle_{0}
\end{eqnarray*}
is invariant under an arbitrary time shift $s$. The proof only requires
the fairly trivial rules $Q^{2}=Q$, $\left\langle XQY\right\rangle _{0}=\left\langle \left(QX\right)Y\right\rangle _{0}$
and $\left\langle XLY\right\rangle _{0}=-\left\langle \left(LX\right)Y\right\rangle _{0}$
to move operators from a factor $X$ to a factor $Y$.

Time inversion symmetry then shows that the correlation function is
a symmetric matrix\footnote{As usual, if external magnetic fields are important then time inversion
symmetry also requires to invert the external currents generating
the magnetic fields.} 
\[
\chi_{i,j}\left(t\right)=\Bigl\langle R_{i}\left(0\right)R_{j}\left(t\right)\Bigr\rangle_{0}=\Bigl\langle R_{i}\left(0\right)R_{j}\left(-t\right)\Bigr\rangle_{0}=\Bigl\langle R_{i}\left(t\right)R_{j}\left(0\right)\Bigr\rangle_{0}=\chi_{j,i}\left(t\right).
\]
Because $R$ is a fast variable $\chi_{i,j}\left(t\right)$ decays
relatively fast, in a time of order $\tau$. But it is not necessarily
true that this decay can be described by a process independent of
$A$. The variables $A$ are quasi constant in a time interval $\tau,$
and the fast variables effectively are restricted to a smaller phase
space. A more detailed equation which takes this dependence into account is
\begin{eqnarray*}
\chi_{i,j}\left(t\right) & = & \int dap_{0}\left(a\right)\Lambda_{i,j}\left(a,t\right),\\
\Lambda_{i,j}\left(a,t\right) & = & \frac{1}{p_{0}\left(a\right)}\Bigl\langle\delta\left(a-A\right)R_{i}\left(0\right)R_{j}\left(t\right)\Bigr\rangle_{0}=\Bigl\langle R_{i}\left(0\right)R_{j}\left(t\right);a\Bigr\rangle_{0},
\end{eqnarray*}
where $\Lambda$ is a correlation function in which $H$ and $A$
are constant. The interesting aspect is that this correlation is identical with the correlation function in the damping term (\ref{eq:K_Generic}).

\section{Markov approximation}

Further simplifications are possible if the slow variables really
are slow in comparison to the fast variables. In that case it suggests
itself to replace the unknown fast variables $R\left(\Gamma,t\right)$
with a Gaussian white noise. The memory effects in the damping (\ref{eq:K_Generic})
disappear and there results the Langevin equation

\begin{equation}
\frac{dA_{i}}{dt}  =  \sum_{j}P\left[A_{i},A_{j}\right]\frac{dH_{eff}}{dA_{j}}-\sum_{j}\lambda_{i,j}\left(A\right)\frac{dH_{eff}}{dA_{j}}+k_{B}T\sum_{j}\frac{d\lambda_{i,j}}{dA_{j}}+r_{i}\left(t\right),\label{eq:Langevin_Final}\\
\end{equation}
\begin{equation}
\Bigl\langle r_{i}\left(0\right)r_{j}\left(t\right);A\Bigr\rangle=2k_{B}T\lambda_{i,j}\left(A\right)\delta\left(t\right).       \label{eq:Langevin_Einstein}
\end{equation}
This is the final result. It contains Onsager reciprocity $\lambda_{i,j}=\lambda_{j,i}$
and the Einstein relations (\ref{eq:Langevin_Einstein}). A special feature
of this Langevin equation is the extra term proportional
to $k_{B}T$ on the r.h.s. The formal derivation explicitly allows damping
coefficients depending on the slow variables $A$. The extra term is a consequence.
This unconventional term is of
order $O\left(A^{0}\right)$ for a linear dependence of $\lambda$ on $A$ and
appears to be large, but this is misleading. As can be seen from
\[
\int dAp_{0}\left(A\right)\left(-\sum_{j}\lambda_{i,j}\left(A\right)\frac{dH_{eff}}{dA_{j}}+k_{B}T\sum_{j}\frac{d\lambda_{i,j}\left(A\right)}{dA_{j}}\right)=0
\]
the extra term cannot be seen in isolation and is required to have
$\left\langle dA/dt\right\rangle =0$. Furthermore, as can be seen
from 
\[
\frac{1}{n}\sum\left\langle A_{j}\frac{dH_{eff}}{dA_{j}}\right\rangle =\frac{-k_{B}T}{n}\sum\int daa_{j}\frac{d}{da_{j}}p_{0}\left(a\right) = k_{B}T
\]
the extra term actually is small in comparison to the second term
if the $A$-dependence of $\lambda$ is small. In that case it thus
is justified to use a conventional Langevin equation with constant
damping coefficients and without the extra term.

It also is of interest that the dependence of the damping coefficients
on the slow variables in a way is trivial at the level of the Langevin
equation. Constant coefficients $\lambda^{\left(A\right)}\left(A\right)$
transform to non-constant coefficients $\lambda^{\left(B\right)}\left(B\right)$
in the Langevin equation for variables $B$ which are in (nonlinear)
bijection with $A$. However, the transformation law of the extra
term is rather complicated, see appendix.

\section{Summary and perspective}

This hopefully sufficiently self-contained article has reproduced
the derivation of generalized nonlinear Langevin equations from classical
statistical mechanics. Questions addressed are the stationarity of
the fluctuating forces and the relevance of unconventional extra terms
caused by the divergence of Poisson brackets and nonlinear damping
coefficients.

In general the formalism doesn't leave much to be desired. A possible
caveat is the division of the degrees of freedom according to the
category fast and slow. There rarely is a clear-cut limit and the
division is more or less arbitrary. Repeating the derivation with
a shifted limit generates a Langevin equation with modified parameters.
This is equivalent to the renormalization group, and the question
is whether the formalism generates a simple Langevin equation of definite
form in the end. In the case of critical phenomena this definitely
is the case, the fixed points of the renormalization group are of
the Langevin equation type, albeit with nontrivial coefficients. The
formalism, however, in principle works with any projection operator,
but interpretation and application of the generalized Langevin equations
then may be difficult.

\section*{Appendix}

\subsection*{Divergence of Poisson brackets}

The mode coupling (\ref{eq:ModeCoupling_V}) contains a term reproducing
the classical mechanics of $A$ with a Hamiltonian $H_{eff}\left(A\right)$
but also an additional purely kinematic term containing the Poisson
bracket divergence 
\[
D_{i}\left(\Gamma,t\right)=-k_{B}T\sum\frac{d}{dA}_{j}P\left[A_{i},A_{j}\right].
\]
The meaning of this extra contribution can be elucidated by repeating
the derivation of the mode coupling term for slow variables $B$ which
are in bijection with $A$. The set of slow variables effectively
is the same and the projection operator is unchanged.\footnote{A special case of this is $P=1$ where $A$ and $B$ are in bijection
with $\Gamma$, without any fast variables.} The new aspect is that now the probability distribution
\begin{equation}
p_{0}^{\left(B\right)}\left(B\right)=const\times exp\left(-\beta H_{eff}^{\left(B\right)}\left(B\right)\right)=p_{0}\left(A\right)\left|dA/dB\right|\label{eq:p0A_p0B}
\end{equation}
of the variables $B$ can be related to the probability distribution
of the variables $A$. This means that $H_{eff}\left(A\right)$ and
$H_{eff}^{\left(B\right)}\left(B\right)$ cannot agree if the Jacobian
isn't constant.

Theorem: If $A$ and $B$ are in bijection then the following three
statements are equivalent:
\[
\forall_{i}\;\sum\frac{d}{dA}_{j}\left[A_{i},A_{j}\right]=0\;\;\;\Leftrightarrow\;\;\;\forall_{G\left(A\right)}\;\sum_{j}\frac{d}{dA_{j}}\left[G,A_{j}\right]=0\;\;\;\Leftrightarrow\;\;\;\forall_{i}\;\sum_{j}\frac{d}{dB_{j}}\left[B_{i},B_{j}\right]\left|\frac{dA}{dB}\right|=0.
\]
The connection between first and second statement is trivial. The
geometric meaning is that every $G\left(A\right)$ generates a divergenceless
Hamiltonian flow in $A$-space. The first statement implies that the
mode coupling for $A$ is of standard form ($D=0$) and that the mode
coupling for $B$ also can be written in a simple form,
\begin{eqnarray}
dA_{i}/dt & = & \sum_{j}\left[A_{i},A_{j}\right]dH_{eff}\left(A\right)/dA_{j},\label{eq:ModeCoupling_B}\\
dB_{i}/dt & = & \sum_{j}\left[B_{i},B_{j}\right]dH_{eff}\left(A\left(B\right)\right)/dB_{j}.\nonumber 
\end{eqnarray}
On the other hand, the formal derivation of an equation of motion
for $B$ as above gives the mode coupling
\begin{eqnarray*}
dB_{i}/dt & = & -k_{B}T\frac{1}{p_{0}^{\left(B\right)}\left(B\right)}\sum_{j}\frac{d}{dB_{j}}\left[B_{i},B_{j}\right]p_{0}^{\left(B\right)}\left(B\right)\\
 & = & \sum_{j}\left[B_{i},B_{j}\right]\frac{dH_{eff}\left(A\left(B\right)\right)}{dB_{j}}-k_{B}T\sum_{j}\frac{d}{dB_{j}}\left[B_{i},B_{j}\right]\left|\frac{dA}{dB}\right|,
\end{eqnarray*}
where (\ref{eq:p0A_p0B}) was used. Comparison with the mode coupling
(\ref{eq:ModeCoupling_B}) deduced from the equation of motion for
$A$ proves the third statement. Equivalence with the first statement
follows by symmetry.

The meaning of all this is that the unconventional extra term $D$
is associated with the Jacobian of the variable transformation.
If the Jacobian is constant and $D\left(A\right)=0$ then also $D\left(B\right)=0$.
In other words, the extra term arises when the exponent of the probability
distribution isn't the energy.

The fact that $D=0$ in standard cases can be understood from
another point of view. Often the variables $A$ directly are related
to the generators of a Lie algebra. For example, for Heisenberg ferromagnets
(model '$J$' of critical dynamics) $A=\left\{ m_{1},m_{2},m_{3}\right\} $,
where the $m_{i}$ are the components of the magnetization density.
The Poisson brackets then generate the Lie algebra (decorated with a $\delta$-function)
\[
\left[A_{i}\left(\mathbf{x}\right),A_{j}\left(\mathbf{x}'\right)\right]=\sum_{m}h_{j,m}^{\left(i\right)}A_{m}\left(x\right)\delta\left(\mathbf{x}-\mathbf{x}'\right),
\]
where the structure constants $h$ are a matrix representation of
the algebra (the adjoint representation). According to a theorem of
group theory real representations of finite and compact groups are
equivalent to orthogonal representations and thus have vanishing trace.
This in part explains why standard Langevin equations don't have the
unconventional extra contribution.

\subsection*{Variable transformations in Langevin equations}

Under a bijection of variables $A\leftrightarrow B$ one has
\[
\Delta B_{i}=\sum\frac{dB_{i}}{dA_{m}}\Delta A_{m}+\frac{1}{2}\sum\frac{d^{2}B_{i}}{dA_{m}dA_{n}}\Delta A_{m}\Delta A_{n}+....
\]
Dividing by a time interval $\Delta t$ and inserting the Langevin
equation (\ref{eq:Langevin_Final}) leads to the Langevin equation
for $B$,
\[
\frac{dB_{i}}{dt}=\sum\frac{dB_{i}}{dA_{m}}\cdot\left(\frac{dA_{m}}{dt}\right)+k_{B}T\sum\frac{d^{2}B_{i}}{dA_{m}dA_{n}}\lambda_{m,n}^{\left(A\right)}.
\]
A word of caution is in order here: this is not standard analysis!
$B$ has to be expanded to second order in $A$ because the product
of two fluctuating forces from the r.h.s. of the Langevin equation
for $A$ is a delta function, see eq.(\ref{eq:Langevin_Final}).

This should agree with the Langevin equation for $B$ a la eq.(\ref{eq:Langevin_Final})
derived directly. There follow the identifications
\begin{eqnarray*}
r_{i}^{\left(B\right)}\left(t\right) & = & \sum_{m}\frac{dB_{i}}{dA_{m}}r_{m}^{\left(A\right)}\left(t\right),\\
\lambda_{i,j}^{\left(B\right)} & = & \sum_{m,n}\frac{dB_{i}}{dA_{m}}\frac{dB_{j}}{dA_{n}}\lambda_{m,n}^{\left(A\right)}.
\end{eqnarray*}
The transformation law of the extra term can be found with the divergence
rule
\[
\sum_{j}\frac{d}{dB_{j}}q_{j}^{\left(B\right)}=\left|\frac{dA}{dB}\right|\sum\frac{d}{dA_{n}}\frac{dA_{n}}{dB_{j}}q_{j}^{\left(B\right)}\left|\frac{dB}{dA}\right|
\]
according to
\begin{eqnarray*}
\sum_{j}\frac{d\lambda_{i,j}^{\left(B\right)}}{dB_{j}} & = & \left|\frac{dA}{dB}\right|\sum\frac{d}{dA_{n}}\frac{dA_{n}}{dB_{j}}\lambda_{i,j}^{\left(B\right)}\left|\frac{dB}{dA}\right|=\left|\frac{dA}{dB}\right|\sum\frac{d}{dA_{n}}\frac{dB_{i}}{dA_{m}}\lambda_{m,n}^{\left(A\right)}\left|\frac{dB}{dA}\right|\\
 & =\sum_{m,n} & \frac{dB_{i}}{dA_{m}}\left(\frac{d\lambda_{m,n}^{\left(A\right)}}{dA_{n}}+\lambda_{m,n}^{\left(A\right)}\frac{d}{dA_{n}}\ln\left|\frac{dB}{dA}\right|\right)+\sum_{m,n}\frac{d^{2}B_{i}}{dA_{m}dA_{n}}\lambda_{m,n}^{\left(A\right)}.
\end{eqnarray*}
Even if the original damping coeffcients $\lambda^{\left(A\right)}$ are
constant then a nonlinear variable transformation generates non-constant
damping coefficients $\lambda^{\left(B\right)}$. However, slow variables normally have a certain meaning
(conserved quantities, order parameter, ...), and a nonlinear transformation thus is rather formal.

%\bigskip{}
%\bigskip{}

\bigskip{}
\bigskip{}
\noindent\textit{Current address:}
\noindent\textsc{Rohde \& Schwarz GmbH \& Co KG, M\"{u}hldorfstr. 15, 81671 Munich, P.O.B. 801469.}
\par\nopagebreak
\noindent\textit{E-mail:} \texttt{rdengler@cablemail.de}


\begin{thebibliography}{1}
\bibitem{Zwan1961}R. Zwanzig, Phys. Rev. 124 (1961) 983. \textquotedbl{}Memory
Effects in Irreversible Thermodynamics\textquotedbl{}

\bibitem{Kaw1973}K. Kawasaki 1973 J. Phys. A: Math. Nucl. Gen. 6 1289.
``Simple derivations of generalized linear and nonlinear Langevin equations''

\bibitem{Mori1965}H. Mori, Prog. Theor. Phys. 33 (1965) 423. \textquotedbl{}Transport,
Collective Motion, and Brownian Motion\textquotedbl{}
\end{thebibliography}
\end{document}